\documentclass[twocolumn,runningheads]{svjour2}

\smartqed  % flush right qed marks, e.g. at end of proof
\usepackage{graphicx}
\newcommand{\g}{$\gamma$}
\newcommand{\n}{$\nu~$}
\newcommand{\e}{\epsilon}

\newcommand{\gp}{\gamma_p}
\newcommand{\ep}{\epsilon^\prime}
\newcommand{\lesssim}{\lower.5ex\hbox{$\; \buildrel < \over\sim \;$}}
\newcommand{\gtrsim}{\lower.5ex\hbox{$\; \buildrel > \over\sim \;$}}

\journalname{Astrophysics and Space Science}
\begin{document}

\title{High-Energy Cosmology}
\subtitle{\g~rays and neutrinos from beyond the galaxy
}


\author{Charles D.\ Dermer
}


\institute{C.~Dermer \at
              Code 7653, Naval Research Laboratory \\
		  4555 Overlook Ave., SW\\
		  Washington, D.\ C.\ 20375-5352 USA\\
              Tel.: +1-202-767-2965\\
              Fax: +1-202-767-0497\\
              \email{dermer@gamma.nrl.navy.mil}           %  \\
}

\date{Received: date / Accepted: date}

\maketitle

\begin{abstract}
Our knowledge of the high-energy universe is undergoing a period
of rapid change as new astronomical
detectors of high-energy radiation start to operate
at their design sensitivities. Now is a boomtime for 
high-energy astrophysics, with new discoveries from Swift and HESS,
 results from MAGIC and VERITAS starting to be 
reported, the upcoming launches of the
 \g-ray space telescopes GLAST and AGILE, and anticipated 
data releases from IceCube and Auger. 

A formalism for calculating statistical properties
of cosmological \g-ray sources is presented. 
Application is made to model calculations
of the statistical distributions of  \g-ray and neutrino emission from 
($i$) beamed sources, specifically, long-duration GRBs, blazars, and extagalactic
microquasars, and ($ii$) unbeamed sources, including normal galaxies, starburst galaxies
and clusters. Expressions for the integrated intensities
 of faint beamed and unbeamed high-energy radiation 
sources are also derived. A toy 
model for the background intensity of radiation
from dark-matter annihilation
taking place in the early universe is constructed. Estimates for the $\gamma$-ray fluxes
of local group galaxies, starburst, and infrared luminous galaxies are briefly reviewed.
 
Because the brightest extragalactic
\g-ray sources are flaring sources, and these are 
the best targets for sources of PeV -- EeV neutrinos and ultra-high
energy cosmic rays, rapidly slewing all-sky telescopes like MAGIC
and an all-sky  \g-ray observatory beyond Milagro will be crucial for optimal
science return in the multi-messenger age.

\end{abstract}

\section{Introduction}
\label{intro}

The next decade is likely to be remembered as the pioneering epoch when the first 
high-energy (PeV -- EeV) $\nu$ sources were detected
with IceCube \cite{hal06} and its km-scale Northerm hemisphere counterpart, 
and when the problem of cosmic-ray origin
was finally solved through identification of the sources of 
cosmic rays at all energies, from GeV -- TeV nucleonic cosmic rays accelerated
by supernova remnant shocks of various types, to 
extragalactic super-GZK \g-ray and $\nu$ sources. 
% The 
%nature and origins of the unidentified EGRET source populations will
%be discovered. Statistical analyses of 
%$\gamma$-ray bursts (GRBs) and blazar AGNs will, after deciphering
%the information encoded in the flaring phenomenology, reveal the
%star formation history of GRB progenitors and the history of
%supermassive black hole assembly, respectively.

The cosmology of  $\gamma$-ray sources in the 
$\approx 10$ MeV -- 10 GeV range  is treated here. 
The lower bound of this energy range
ensures that the $\gamma$ rays
originate from nonthermal processes, and the upper bound 
is defined by the energies of photons that originate from sources
at redshifts $z \gg 1$ without significant 
$\gamma\gamma \rightarrow e^+ e^-$ attenuation in reactions with photons
of the extragalactic background light (EBL). 
The formalism also applies to other nonthermal radiations,
to ultra-relativistic particles, including
PeV -- EeV $\nu$ and ultra-high energy neutrals, and to multi-GeV
-- EeV photons by taking into account attenuation and reprocessing of
the $\gamma$-rays on the EBL. 

The problems treated here are the
\begin{enumerate}
\item Event rate of bursting sources;
\item Size distribution of bursting sources; and 
\item Apparently diffuse intenstity from unresolved sources.
\end{enumerate}
I outline applications of these results to beamed souces, including
GRBs, blazars and extragalactic microquasars, and unbeamed
sources, including star-forming galaxies
and merging clusters of galaxies. 

This paper, prepared
for the conference proceedings of the {\it Multi-Messenger Approach to 
High Energy Gamma-Ray Sources}, held 4 -- 7 July 2006 in Barcelona, Spain,
addresses in a more formal manner the points I was to cover, including
blazars which I could not neglect (see Ref.\ \cite{bot06} for 
a review of blazar emissions). The formalism applies to analysis
of $\gamma$-ray and $\nu$ data from GLAST, IceCube,
and other high-energy astroparticle observatories.

\section{Event Rate of Bursting Sources}

The Robertson-Walker metric for a homogeneous, isotro-pic universe
can be written as 
\begin{equation}
ds^2 = c^2dt^2 -R^2(t)\big( {dr^2\over 1-kr^2}+r^2 d\Omega\big)\;,
\label{RW}
\end{equation}
where $r$ is a comoving coordinate and $R(t)$ is the expansion 
scale factor. The most convenient choice is to have $r$ take the value of
physical distance at the present epoch so that $R(t) = R = 1$, 
and denote $R_* = R(t_*)$ at emission time $t_*\leq t$ (stars denote the emission epoch). 
Material structures reside for 
the most part on constant values of the comoving coordinates, 
whereas light and ultra-relativistic particles cannot be confined
to such coordinates. From the definition of redshift 
$z = (\lambda - \lambda_*)/\lambda_*$, we have
$1+z = \epsilon_*/\e = \Delta t/\Delta t_* = R/R_*$, where $\e$ refers
to the energy of the photon or ultrarelativistic particle. The curvature of space
is determined by the curvature constant $k$, with $k = 0$ for flat space.

The proper volume element of a slice of the universe at time $t_*$ is,
from eq.\ (\ref{RW}) for a flat universe, 
\begin{equation}
dV_* =  \; R_*^3 r^2 dr = dr_* r_*^2 d\Omega_*  = cdt_* dA_*\;.
\label{dV*}
\end{equation}
Comparing with the definition
$dA = (Rr)^2 d\Omega$ and noting that $d\Omega_* = d\Omega$ in the 
absences of cosmic shear, we have
\begin{equation}
{dA_*\over dA} =  {1\over (1+z)^2}\;.
\label{dA*}
\end{equation}

The directional event rate, or event rate per sr, is
$${d\dot N\over d\Omega} \;=\;{1\over 4\pi}\int
dV_*\; \dot n_*(z_*)\;|{dt_*\over dt}|\;=\;$$
\begin{equation}
c\int_0^\infty dz \;|{dt_*\over dz}| {(R_*r)^2 \dot n_*(z)\over (1+z)}, 
\label{ndotOmega}
\end{equation}
where the burst emissivity $\dot n_*(z_*)$ gives the 
rate density of events at redshift $z$.
An expression for $(R_*r)^2$ can be derived by recalling the 
relationship between energy flux $\Phi_E$ and luminosity distance
$d_L$, namely
\begin{equation}
{d{\cal E}\over dA dt} = \Phi_E = {L_* \over 4\pi d_L^2} =  {d{\cal
E}_*\over 4\pi d_L^2d t_*}\; = \;{(1+z)^2\over 4\pi d_L^2}
\;\Phi_E dA\;, 
\label{dEdAdt}
\end{equation}
so that with eq.\ (\ref{dA*}),
\begin{equation}
(R_*r)^2 = {d^2_L(z) \over (1+z)^4}\;. 
\label{dLum}
\end{equation}
For a flat $\Lambda$CDM universe,
\begin{equation}
|{dz\over dt_*}| = H_0 (1+z) \sqrt{\Omega_m(1+z)^3 +
\Omega_\Lambda}\; \label{dzdt*}
\end{equation}
\cite{pee93,spe03}, where $H_0 = 72$ km s$^{-1}$ Mpc$^{-1}$,
$\Omega_m =0.27$ and $\Omega_\Lambda = 0.73$ are the ratios of the
energy densities of total mass, including both normal matter and dark matter, and dark
energy, respectively, compared to the critical density for the flat $\Lambda$CDM cosmology
of our universe.

The directional event rate, eq.\ (\ref{ndotOmega}), becomes
\begin{equation}
{d\dot N\over d\Omega} \;=\;
c\int_0^\infty dz \;|{dt_*\over dz}| {d_L^2(z) \dot n_{co}(z)\over (1+z)^2}, 
\label{ndotOmega1}
\end{equation}
after using the relation $\dot n_* = (1+z)^3 \dot n_{co}(z)$
to write the directional event rate in comoving rather
than proper quantities. If separability between the emission 
properties and the rate density of sources can be assumed (a crucial assumption), 
then $\dot n_{co,i}(z) = \dot n_{i} \Sigma_i(z)$, 
where $\Sigma_i(z)$ is the structure formation history
(SFH) of sources of type $i$, 
defined so that $\Sigma_i(z\rightarrow 0) = 1$, and $\dot n_i$ is the local ($z\ll 1$) 
rate density of bursting sources of type $i$ (see Fig.\ 1).

\begin{figure}
\vskip-0.5in
\centering
  \includegraphics[scale=0.45]{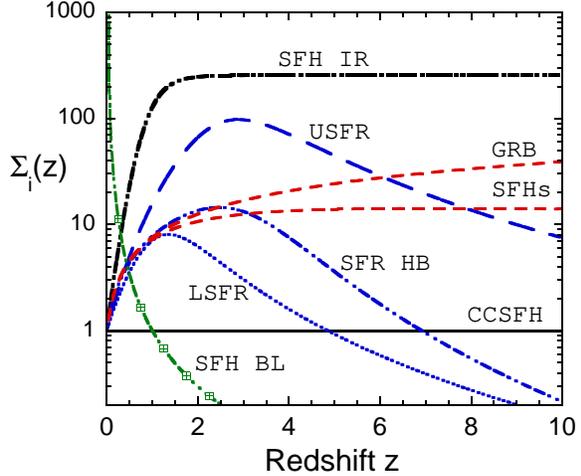}
\vskip-0.1in
\caption{Different structure formation histories (SFHs) considered in this 
paper. As labeled, CCSFH: constant comoving SFH; LSFR: lower star formation rate;
USFR: upper SFR; SFR HB: SFR history from Ref.\ \cite{hb06}; SFH IR: SFH of IR luminous
galaxies \cite{san04}; GRB SFHs: range of SFHs of GRBs used to fit Swift and pre-Swift 
GRB distributions \cite{ld06}; SFH BL: SFH of BL Lac objects \cite{der06}. These rates
are poorly known at $z\gg 1$.}
\label{fig:1}      
\end{figure}

The comoving density can be formally expanded as
\begin{equation}
 \dot n_{co}(z) =\oint d\Omega^\prime \int_0^\infty d\alpha \;N(\alpha;z) 
\int \dots \dot n_{co} (\Omega^\prime,\alpha ,\dots ;z)
\label{ncoz}
\end{equation}
\cite{der06,ld06}.
The direction $\Omega^\prime = (\mu^\prime=\cos\theta^\prime,\phi^\prime)$ 
specifies the orientation of the 
system with respect to the direction to the observer, and 
$N(\alpha;z)$ is a normalized distribution function for
parameter $\alpha$. For example, $\alpha$ could represent the 
bulk Lorentz factor $\Gamma$, the total energy radiated, the comoving-frame
power, or
the spectral index of the radiation. For sources oriented at random,
$\dot n_{co} (\Omega^\prime,\alpha ,\dots ;z) = \dot n_{co} (\alpha ,\dots ;z)/4\pi$.
For two-sided jet sources, $-1 \leq \mu^\prime \leq 1$ and 
$\dot n_{co} (\Omega^\prime,\alpha ,\dots ;z) = 2\dot n_{co} (\alpha ,\dots ;z)/4\pi$.

If one were to consider persistent rather than bursting
sources, an analogous derivation gives the directional 
number count of sources of the type $i$ as
\begin{equation}
{dN_i\over d\Omega} = cn_i \int_0^\infty dz\; 
|{dt_*\over dz}| {d_L^2(z) \Sigma_{i}(z)\over (1+z)}. 
\label{dNidOmega}
\end{equation}

\section{Size Distribution}

After substituting eq.\ (\ref{ncoz}) into eq.\ (\ref{ndotOmega1}) and 
placing limits on the integrals in accordance with detector specifications,
model distributions of source properties can be derived, in particular,
the size distribution. 

\subsection{Beamed Sources}

A distinction between two types of models for relativistically beamed sources
needs to be made. A {\it blast-wave model} is usually considered for GRB sources. 
Here the half-angular extent $\theta_j$ of the collimated spherical blast-wave jet is 
assumed to be much greater than the 
Doppler beaming angle $\theta_{\rm D} \sim 1/\Gamma$, so that $\theta_j \gg \theta_{\rm D}$, 
where $\Gamma$ is the bulk 
Lorentz factor of the outflow. In this case, the observer is limited to detection of a GRB
if the direction from the source to the observer intercepts the solid angle of the blast wave. 
By contrast, in a {\it blob model}, as is usually considered
in blazar studies,  $\theta_j\ll \theta_{\rm D}$, 
and the beaming properties of the jet are determined
primarily by the Dopper factor 
\begin{equation}
\delta_{\rm D} = [\Gamma(1-\beta\mu)]^{-1}\;,
\label{deltaD}
\end{equation}
where $\beta\Gamma = \sqrt{\Gamma^2 -1}$, and $\arccos\mu$
is the observer's angle measured with respect to the jet 
axis. Interpreting emissions within a blast-wave and blob framework 
provide the simplest models that can be used to systematically analyze
the statistics of GRBs, blazars, radio galaxies, and microquasars 
within the context of a physical (rather than a phenomenological) model.

\paragraph{Gamma Ray Bursts}

If a GRB releases an amount of $\gamma$-ray energy ${\cal E}_{*\gamma}$ that is deposited in 
a waveband to which a GRB detector is sensitive, then an event is recorded when
the source flux 
\begin{equation}
{  {\cal E}_{*\gamma}\over 4\pi d_L^2(z)(1-\mu_j) \Delta t_* \lambda_b} \gtrsim f_\epsilon\;,
\label{Estarg}
\end{equation}
where $f_\e$ is the $\nu F_\nu$ threshold sensitivity of the GRB detector. 
Beaming of the emission into a jet with opening half-angle $\theta_j$ has the 
effect, for constant ${\cal E}_{*\gamma}$, to
enhance the received flux by a factor $(1-\mu_j)^{-1}$, though
the chance of the jet being in the line of sight to the observer
(compared to an isotropically emitting source) is 
reduced 
by the factor $1-\mu_j$. The term $\lambda_b$ is a bolometric correction 
factor made in lieu of a full spectral treatment.

The GRB size distribution for the blast-wave geometry is given by
$$\frac{d\dot{N}_{GRB}(> f_\e)}{d\Omega  }  =  2c \dot n_{GRB} 
\int_0^\infty dz\;  |{dt_*\over dz}| \frac{d^2_L(z)\Sigma_{GRB}(z)
}{(1+z)^2} \;$$
\begin{equation}
\times \int_{\max(0,\hat\mu_j)}^1 d\mu_j\;g(\mu_j)\;(1-\mu_j),
\label{dNdtdO>fe}
\end{equation}
where $g(\mu_j)$ is the normalized distribution function of 
GRB jet opening angles, and 
\begin{equation}
\hat\mu_j =  1-{ {\cal E}_{*\gamma}\over 4\pi d_L^2(z) \Delta t_* \lambda_b f_\epsilon}\;.
\label{hatmuj}
\end{equation}

Le and Dermer (2006) \cite{ld06} have used this approach to analyze 
the redshift and opening-angle ($\theta_j$) distributions of GRB detectors, 
including missions before Swift compared with distributions measured
with Swift. They find that the comoving rate densities of GRBs must undergo positive
evolution to at least $z \gtrsim 5$ -- 7 to account for the difference
in distributions of pre-Swift and Swift-detected GRBs with redshift information. 
By contrast, the  star-formation history of the universe as inferred from 
blue and UV luminosity density, peaks at $z\approx 2$ -- 3
and seems to decline at larger redshift \cite{hb06}. Le and Dermer (2006) 
find this SFH to be
incompatible with the statistics of GRBs with measured redshifts. Thus the SFH
 of GRBs is apparently very different than the integrated 
high-mass star formation history of the universe.

This approach can be suitably adapted to the short, hard class of GRBs to infer
the rate density of this class of GRBs. A large data set, accumulated after a
long Swift lifetime, can in principle distinguish between models 
involving compact-object coalescence and accretion-induced collapse of neutron 
stars.

\paragraph{Blazars}

A considerable simplification to the emission properties of 
blazars results by approximating the 
$\nu F_\nu$ fluxes detected from a distant source
by the expression
\begin{equation}
f_\e^{proc} = {\ell^\prime_e \delta_{\rm D}^q \e_z^{\alpha_\nu}\over d_L^2(z)}\;\geq \; f_\e\;
\label{feproc}
\end{equation}
\cite{da04,der06}, where $\ell^\prime_e$ is the comoving directional power
 and $f_\e$ represents the characteristic flare size, in this case,
in units of energy flux. The beaming factor indices for individual radiating blobs are 
$$ q = \cases{(p+5)/2,\; & synchrotron/SSC $~$ \cr\cr p+3\;
&EC$~$ \cr}\;.\; $$ 
The synchrotron/SSC factor applies to blazars where the $\gamma$ rays are from the 
synchrotron self-Compton processes, specifically X-ray--selected blazars and TeV blazars. The external Compton
(EC) beaming factor applies to blazars where the \g-rays are ambient photons, external to the
jet, that intercept the jet and Compton-scattered by the jet electrons. Examples of ambient radiation 
fields are the accretion disk photons, and accretion-disk photons that are scattered by surrounding dust and gas.

From eqs.\ (\ref{ndotOmega1}) and (\ref{ncoz}), the blazar flare size distribution is given by the expression 
$${d\dot N_{bl}\over d\Omega} (>f_\e )\;=\; 2c \dot n_{bl} \;\int_0^\infty dz\; 
|{dt_*\over dz}|\;{d_L^2(z) \Sigma_{bl}(z)\over (1+z)^2}\times
$$
\begin{equation}
\int_1^\infty d\Gamma\; N(\Gamma;z)\int_0^\infty d\ell^\prime_e\;N(\ell^\prime_e;z)\; [1-\max(-1,\hat\mu)]\;,
\label{ddotNdOf3}
\end{equation}
where 
\begin{equation}
\hat \mu  = {1\over \beta} \big[ 1 - {1\over \Gamma}\big({\ell^\prime_e \e_z^{\alpha_\nu}\over d_L^2 \;f_\e}\big)^{1/q} \big] \;.
\label{hatmu1}
\end{equation}
Specification of the $z$-evolution of the normalized distribution functions
$N(\Gamma;z)$ and $N(\ell^\prime_e;z)$ due to number evolution or luminosity evolution, respectively, 
connects this formulation back to the cosmology of physical processes and the growth of 
structure taking place in the early universe. 

Refs.\ \cite{der06} uses this approach to analyze the 
redshift and size distribution of EGRET \g-ray blazars (see also \cite{mp00}), 
divided into flat spectrum radio quasars (FSRQs) and BL Lac objects (BLs).
Evolutionary behaviors are found that characterize the measured 
redshift and size distributions of FSRQs and BLs.
The behavior of the BLs is in accord with the 
conjecture that BLs 
are late stages of the formation and evolutionary history of FSRQs, and before that,
IR luminous galaxies \cite{bd01,san04}. See
 Ref.\ \cite{der06} for predictions of the number
of blazars that GLAST will detect.

\paragraph{Microquasars}

\g-ray emission from microquasars could be visible from nearby galaxies if
the bulk Lorentz factors in microquasar jets were large enough that the received
flux from a microquasar in another galaxy was brighter than threshold. 
The size distribution of microquasar flares
can be written by taking the limit $z\ll 1$ of  the blazar expression, eq.\ (\ref{ddotNdOf3}),
to give
 $${d\dot N_{\mu q}\over d\Omega} (>f_\e )\;=\; {2c^3 \dot n_{\mu q} \over H_0^3}\;\int_0^\infty dz\; 
z^2 \Sigma_{\mu q}(z)
$$
\begin{equation}
\times \int_1^\infty d\Gamma\; N(\Gamma;z)\; [1-\max(-1,\hat\mu)]\;,
\label{ddotNdOf4}
\end{equation}
which assumes an averaging over the small scale mass distributions of nearby galaxies. 
The integration in $\ell_e^\prime$ is removed in this expression,
 compared with eq.\ (\ref{ddotNdOf3}), by assuming an
Eddington limitation on the accretion flow. Approximating the 
emission spectrum by a single power law with $\nu F_\nu$ index $\alpha_\nu$
in the comoving energy range $\ep_0 < \ep < \ep_1$,
 the directional luminosity is therefore
limited by 
\begin{equation}
%\int_{\ep_0}^{\ep_u} d\ep\; \ell^\prime_e \e^{\prime \alpha_\nu - 1} 
\ell^\prime_e\lesssim {2\times 10^{38} m_{\rm C}\over 4\pi\lambda_b}\; {\rm~ergs~s}^{-1}\;{\rm sr}^{-1}\;,
\label{intep0}
\end{equation}
noting that the emission is beamed into  
$\approx \Gamma^{-2}$ of the full sky, and that the radiated power
is boosted by $\Gamma^2$ due to bulk motion of the plasma.
Here $m_{\rm C}$ is the Chandrasekhar mass (in units of 1.4 $M_\odot$)
of the compact object in the microquasar. 
%Thus $\ell^\prime_e \lesssim 
%2\times 10^{38} m_{\rm C}/ 4\pi\lambda_b$.

\subsection{Unbeamed Sources}

For $\gamma$-ray emission from unbeamed sources, like the Milky Way
galaxy, normal galaxies, and all but the most dusty and 
heavily extincted starburst and infrared luminous galaxies (whose
ambient radiation would attenuate the $\gamma$ rays), we can 
count the number of source detections above a threshold flux 
$f_\e$, following eq.\ (\ref{dNidOmega}), to give:
$${dN_{i}\over d\Omega} (>f_\e )\;=\;c  n_{i} \;\int_0^\infty dz\; 
|{dt_*\over dz}|\;{d_L^2(z) \Sigma_{i}(z)\over (1+z)}$$
\begin{equation}
 \times\int_{L_{*min}}^\infty dL_*\; N(L_*;z)\;.
\label{ddotNi}
\end{equation}
The luminosity function of the unbeamed source population 
is denoted by $N(L_*;z)$, where $L_* = \int_0^\infty d\e_* L(\e_*)$
is the total luminosity of the source. Writing the spectral luminosity 
$L_*(\e_*) = L_{*0} \e_*^{-1+\alpha_\nu}$ gives the 
threshold condition 
\begin{equation}
{L_{*0}\e_*^{\alpha_\nu} \over 4\pi d_L^2 } \geq f_\e\;
\label{L*0e}
\end{equation}
for detection of these sources. For a power-law spectrum with 
low- and high-energy cutoffs, this expression can be used to 
impose the lower limit $L_{*min}$ in eq.\ (\ref{ddotNi}), 
which also assumes an average over large volumes. 
For normal galaxies, volumes of radii of several Mpc may be large enough for 
this averaging.
For clusters of galaxies, an averaging size scale of many 
tens of Mpc is needed, as calculations at scales less than $z\sim 0.02$ are 
subject to strong fluctuations due to the low density of clusters
of galaxies in this volume.

\section{Intensity of Unresolved Sources}

The differential spectral flux
\begin{equation}
d\phi(\e )= {dN\over dAdt d\e }= {\dot n_*(\e_*;z)d\e_*dt_*dV_*
\over dAdtd\e } \;.
\label{dphie}
\end{equation}
Using the relations $dV_* = dr_* dA_* = cdt_* dA/(1+z)^2$ from 
eq.\ (\ref{dA*}), 
$\e_* = \e(1+z) \equiv \e_z$, and $dt = dt_*( 1+z)$, we have
\begin{equation}
\phi(\e )= c\int_0^\infty dz\; |{dt_*\over dz}|{\dot n_{*}(\e_*;z)
\over (1+z)^2}\;.\;
\label{phie}
\end{equation}
Because the ``$\nu F_\nu$" intensity $\e I_\e = m_ec^2 \e^2 \phi(\e)/4\pi$,
\begin{equation}
\e I_\e = {c\over 4\pi}\int_0^\infty dz\; |{dt_*\over dz}|\;{m_ec^2 \e_*^2 \dot n_{co}(\e_*;z)
\over 1+z}\;.\;
\label{eIe}
\end{equation}

\paragraph{GRBs}

The diffuse intensity of GRBs is, from eq.\ (\ref{eIe}) and 
assuming a two-sided GRB jet source,
$$\e I^{GRB}_\e (<f_\e) = {m_ec^3\dot n_{GRB}
\over 4\pi}\int_0^\infty dz\; |{dt_*\over dz}|\;{ \Sigma_{GRB}(z)
\over 1+z}$$
\begin{equation}
\times \;\int_{0}^{\min(1,\hat\mu_j)} d\mu_j\; g(\mu_j ) (1-\mu_j)\; \e_*^2 N(\e_*;\mu_j).\;
\label{eIeGRB}
\end{equation}
For a flat $\nu F_\nu$ spectrum that covers the waveband of 
the GRB detector, $m_ec^2 \e_*^2 N(\e_*;\mu_j) = {\cal E}_{*\gamma}/[\lambda_b (1-\mu_j)]$,
and $\hat\mu_j$ is given by eq.\ (\ref{hatmuj}).

Using the parameters derived from analyses of statistical 
distributions of GRB data \cite{ld06}, one can then 
calculate the integrated $\gamma$-ray background from GRBs 
which, as we shall see, is a negligible fraction of the diffuse
isotropic \g-ray background. Suitable scalings are adopted
in model calculations of \n-emissions from GRBs to calculate
the diffuse $\sim 100$ TeV -- EeV \n intensity from GRBs \cite{mn06}. 

\paragraph{Blazars}

The total intensity from two-sided blazar jet sources, which 
will include emission from aligned and misaligned blazars and radio galaxies, 
is given in the blob framework by
$$\e I^{bl}_\e = {c\over 2\pi }\;\int_0^\infty dz\; |{dt_*\over dz}|{1\over 1+z}$$
\begin{equation}
\times \oint d\Omega^\prime
\e_*^2 q_{bl}(\e_*,\Omega^\prime;z)
\;,
\label{eIebl}
\end{equation}
 where $q_{bl}(\e_*,\Omega^\prime;z)$ is the directional spectral flux 
of a blazar jet, given in the blob framework by 
$$\e_*^2 q_{bl}(\e_*,\Omega^\prime;z) = \ell_e^\prime(z) n_{bl}(z) \delta_{\rm D}^q \e_z^{\alpha_\nu}$$
\cite{der06}. Here $n_{bl}(z)$ is the comoving density of blazar sources.
The intensity of unresolved blazars and radio galaxies
is then
$$\e I^{bl}_\e (<f_\e )\hskip-0.025in =\hskip-0.025in {c\e^{\alpha_\nu}\over \beta\Gamma^q }\int_0^\infty dz |{dt_*\over dz}|
{n_{bl}(z)\over (1+z)^{1-\alpha_\nu}}\hskip-0.05in\int_1^\infty d\Gamma N(\Gamma;z)$$
\begin{equation}
\times\int_0^\infty d\ell_e^\prime N(\ell_e^\prime;z)
\{[1-\beta \min(1,\hat\mu)]^{1-q} - (1+\beta)^{1-q}\}\;,\;
\label{eIebl1}
\end{equation}
with $\hat\mu$ given by eq.\ (\ref{hatmu1}).

\paragraph{Microquasars}

The intensity from microquasars is given essentially by eq.\ (\ref{eIebl}), though 
with a very different local rate density $\dot n_{\mu q}$, SFH $\Sigma_{\mu q}$, 
and distribution in $\Gamma$ and $\ell_e^\prime$.
 
\paragraph{Unbeamed Sources}

If $N(\Gamma_*;z)$ is the redshift-dependent luminosity function of 
unbeamed $\gamma$-ray sources, such as normal galaxies and starburst
and IR luminous galaxies, then the diffuse intensity from these sources
over cosmic time is 
$$\e I_\e^{iso}(<f_\e) = {c\over 4\pi }\;\int_0^\infty dz\; |{dt_*\over dz}|\;{m_ec^2 \e_*^2 \dot n_{iso}(\e_*;z)
\over 1+z}$$
\begin{equation}
\times \int_0^{L_*^{min}(z)} dL_*\; N(L_*;z)\;,
\label{eIeiso}
\end{equation}
where now  $L_*^{min}(z)$ again depends on detector 
characteristics according to the prescription of eq.\ (\ref{L*0e}).

It is interesting to note that the factor $|dt_*/dz|$ associated
with the passage of time in an expanding universe saves us from Olbers' paradox.
In this formulation, the logarithmically divergent integrated intensity emitted by radiant sources
distributed uniformly throughout the universe is blocked by the redshifting of radiation
and the finite age of the universe.

\paragraph{GZK Neutrino Intensity}

The intensity of \n formed as photopion secondaries in the interaction 
of UHECRs with the EBL is given, starting with eq.\ (\ref{eIe}), in the form
\begin{equation}
\e I^{GZK}_\e = m_ec^3\e^2\int_0^\infty dz\; |{dt_*\over dz}|\;{ \dot n_{GZK,*}(\e_*,\Omega_*;z)
\over (1+z)^2}\;.\;
\label{eIeGZK}
\end{equation}
The production spectrum of secondary \n is given by 
$$\dot n^*_{GZK}(\e_*,\Omega_*;z) = c\sum_{j} \oint d\Omega \;\int_0^\infty d\e_* \;n^*_{ph}(\e_*,\Omega;z) $$
\begin{equation}
\times\oint d\Omega^*_{p}\int_1^\infty d\gp^*\;(1-\cos\psi)\;n_p^*(\gp^*,\Omega_p^*;z)\;
{d\sigma_j(\e^\prime)\over d\e_* d\Omega_*}.\;
\label{eIeGZK1}
\end{equation}
The sum is over various channels leading to production of neutrinos, 
and $n^*_{ph}(\e_*,\Omega;z)$ and $n_p^*(\gp^*,\Omega_p^*;z)$ are the evolving
EBL and UHECR proton spectra, respectively (generalization to ions is straightforward).
Ref.\ \cite{dh06} uses this formalism to calculate the GZK \n 
intensity under the assumption that the sources of UHECRs are 
GRBs. 

The GZK \g-ray intensity can be calculated according to this
formalism by convolving the redshift-dependent differential 
intensity with a source function that represents the emergent
\g-ray spectrum after reprocessing on the background radiation
field. This will produce a complete model of UHECRs, which consists
of a fit to the UHECR spectrum, a prediction for the GZK \n  
flux, and the predicted diffuse \g-ray spectrum---which 
must be less than the diffuse EBL at \g-ray energies \cite{sig06}.

\paragraph{Dark Matter Annihilation}

Astrophysical searches for signatures 
of dark matter annihilation target regions of enhanced 
(dark) matter density, such as cuspy cores of quasi-spherical
galaxies, for example, dwarf ellipticals. Because of its proximity,
even the center of the Milky Way is considered to be 
a hopeful site of dark matter annihilation, in spite of 
perturbing warps and bars in its normal matter distribution.
One region where unavoidably high densities of dark matter
had to persist was in the early universe.

Up to now, we have considered classes of sources whose
density scales as a redshift-dependent structure 
formation rate $\Sigma_i(z)$ for sources of type $i$.
To first approximation, source density is proportional
to the total normal matter content, so that the factor 
$(1+z)^3$ is removed and the SFH is described in terms
of the comoving rate density. Dark matter annihilation
scales as the square of the density of matter, so 
to first order we can write the diffuse background intensity
from dark matter annihilation as 
\begin{equation}
\e I_\e^{DM} = {m_ec^3\e^2\over 4\pi }\; \int_0^{z_{max}} dz\; |{dt_*\over dz}|
\; { \dot n_*^{DM}(\e_*;z)\over (1+z)^2}\;.
\label{eiedm}
\end{equation}
The spectral production rate of dark matter secondaries
is written as
\begin{equation}
\dot n_*^{DM} \simeq j_\chi n_{DM}^2 (1+z)^6 \sigma_{DM}\delta(\e_*-\e_\chi)\;,
\label{dotndm}
\end{equation}
where $\e_\chi$ is the energy of  secondary \g~ rays or \n 
produced in dark matter annihilation, and $j_\chi$ is the multiplicity
of the secondaries. The maximum redshift $z_{max}$ represents 
the redshift where dark matter
was created or fell out of equilibrium.

Hence 
\begin{equation}
\e I_\e^{DM} \propto \int_0^{z_{max}} dz
\; {\e^2 \delta[\e(1+z) - \e_\chi]\over (1+z)^{5/2}}\;\propto (\e/\e_\chi)^{-1/2}
\label{eiedm2}
\end{equation}
for $\e \gtrsim \e_\chi/z_{max}$. A diffuse \n background from 
annihilation of dark matter particles with masses of $\sim 10$ GeV -- TeV
could peak near 10 -- 100 MeV, for $z_{max} \sim 10^4$. A component of
the diffuse extragalactic
$\gamma$-ray background in the 10 MeV -- GeV range would be formed under
the same circumstances as the annihilation $\gamma$ rays cascaded to 
photon energies where the universe becomes transparent to $\gamma\gamma$
pair production. 

If $z_{max}$ corresponds, however, to a redshift where the temperature of the CMB
corresponds to the dark matter particle energy, then this indirect 
signature of dark matter annihilation is probably not detectable.
In any case, a residual photonic or $\nu$ 
signature from dark matter annihilation in the
early universe will be associated with any assumed dark matter 
annihilation cross section, and this emission 
signature cannot exceed measured values or upper limits.

%%%%%%%%%%%%%%%%%%%%%%%%%%%%%%%%%%%%%%%%%%%%
\section{Discussion}
%%%%%%%%%%%%%%%%%%%%%%%%%%%%%%%%%%%%%%%%%%%%

By inferring source densities and event rates from 
astronomical observations (Fig.\ 1), fits can be made to 
statistical (e.g., redshift and size) distributions
 of high-energy sources detected
with GLAST and other high-energy telescopes.  
A model that fits the distributions entails an imperative, for each source class, 
to show that the superpositions of radiations formed by the faint 
model sources below the detection limit
do not overproduce the measured diffuse $\gamma$-ray background 
and upper limits to the \n background radiations. We illustrate
the technique in what follows, first considering detection 
of quasi-isotropic cosmic-ray induced emissions from 
star-forming galaxies.

\subsection{\g-rays and \n from Unbeamed Sources}

Cosmic-ray 
induced emissions from extragalactic sources
are weak \cite{pf01,tor04}, which must be the case in order to 
agree with the lack of high significance 
detections of \g~rays from star-forming galaxies without jets. 
Other than the LMC \cite{sre92}, which was observed with EGRET
at a flux level consistent with that expected if cosmic rays
were produced at a rate proportional to the star formation rate of the Milky Way, 
no unbeamed extragalactic high-energy radiation source has yet
been detected with high confidence. 

\paragraph {Normal, Starburst, and Infrared Luminous Galaxies.}

If cosmic-ray induced emissions are primarily responsible 
for the high-energy quasi-omni directional emissions from extragalactic 
sources, then the level of emission from the Milky Way can 
be appropriately scaled to estimate the expected flux levels
of galaxies of different types. The \g-ray photon production 
rate from the Milky Way inferred from COS-B observations \cite{blo84} 
is 
$\dot N_\gamma \approx (1.3 - 2.5)\times 10^{42}$ ph($>$100 MeV) s$^{-1}$
implying a $> 100$ MeV \g-ray luminosity from the Milky 
Way of $10^{39}L_{39}$ ergs s$^{-1}$, with
$L_{39} =(0.16 - 0.32)$. 
Analysis using GALPROP Galactic cosmic-ray propagation model
\cite{smr00} indicates that, taking into account the GeV excess in the diffuse
galactic emission observed with EGRET and using a 
larger model Milky Way halo,  the $> 100$ MeV \g-ray luminosity of the Milky Way 
is $10^{39}$ ergs s$^{-1}$, with $L_{39} = 
(0.71 - 0.92)$. Some 90\% 
of this emission is due to secondary nuclear production when 
cosmic rays collide with gas and dust in the Galaxy.

Approximating the integrated $> 100$ MeV 
photon spectrum as a power law with a 
mean photon spectral index = 2.4 implies that
the $\nu L_\nu$ spectrum of the Milky Way is
\begin{equation}
\e L_{MW}(\e ) \cong 3.3\times 10^{39} L_{39}\e^{-0.4}{\rm~ergs~s}^{-1}
\label{estarLMW}
\end{equation}
for $\e = h\nu/m_e c^2 \gtrsim 200$ (i.e., $> 100$ MeV),
or $\dot N(> 100$ MeV)$\cong 10^{43}$ ph s$^{-1}$.

By scaling nearby galaxies according to their supernova
rates, a simple estimate for the \g-ray and 
\n~emissions can be made. For example, the supernova
rate of Andromeda (M31),  at a distance of $\approx 800$ kpc, is $\approx
1$ per century, compared to the rate of 2.5 every century
in the Milky Way \cite{pf01}.  Thus the expected \g-ray photon flux 
from M31 should be at the level 
$$\phi_{M31}(> 100{\rm~ MeV}) \approx$$
\begin{equation}
{1\over 2.5}\;{10^{43}{\rm~ph~s}^{-1}\over 4\pi (800{\rm~kpc})^2} 
\approx 0.9\times 10^{-8}L_{39}~{\rm cm}^{-2}~{\rm s}^{-1}\;, 
\label{M31}
\end{equation}
for $\alpha_\nu = -0.4$, which would be significantly detected with GLAST. 
Pavlidou and Fields (2001) \cite{pf01} perform a more detailed
treatment of local group galaxies and predict that the 
$> 100$ MeV integral photon flux from the SMC, M31, and M33 
are at the levels of $1.7\times 10^{-8}$, $1.0\times 10^{-8}$,
and $0.11\times 10^{-8}$ ph ($> 100$ MeV) cm$^{-2}$ s$^{-1}$, 
respectively. GLAST should therefore detect at least the SMC and M31, 
though its 1 year sensitivity of $\approx 0.4\times 10^{-8}$ ph($>100$ MeV)
cm$^{-2}$ s$^{-1}$ means that few other local group galaxies are
likely to be detected.
The difference between the simple-minded treatment presented
here and their more detailed treatment is a consideration of the total target
mass density of the different galaxies, and diffusion and escape
of cosmic rays from the galaxy, which is especially important for 
the Magellanic Clouds.

By extrapolating the integrated diffuse galactic continuum emission
from the Milky Way to TeV energies, the integral number flux
of \g~rays 
from a Milky-Way type galaxy at the distance $d$ 
is $$\phi(>\e )\cong 2.4\times10^{-5} \eta L_{39}\e^{-1.4}/d({\rm Mpc})^2$$
\begin{equation}
\approx 
{2\times 10^{-13}\eta L_{39}\over (d/{\rm 1~Mpc})^2[E_\gamma({\rm 300~GeV})]^{1.4}}
\;{\rm ph(>E_\gamma )~cm^{-2}~s^{-1}},
\label{vFv}
\end{equation} 
where the $\eta$ factor accounts for the different supernova rates
and target densities for the galaxy under consideration, as well as
the reduced number of \g~rays if the spectrum softens with energy. 
Because the imaging atmospheric Cherenkov telescopes HESS and 
VERITAS have sensitivities of $\approx 4\times 10^{-13}$
ph($> 300$ GeV) cm$^{-2}$ s$^{-1}$ 
for $\sim 50$ hour observations \cite{wee02} (see Fig.\ 2), M31, 
a northern hemisphere source (+41$^\circ$ declination), could be marginally
detectable with VERITAS in long exposures, provided that the spectral GeV-TeV softening and 
reduction in sensitivity due to M31's  angular extent, $\sim 1^\circ$, are not too great. 

The enhanced supernova rate in starburst galaxies such as
M82 and NGC 253 at $\approx 3$ Mpc improve the prospects that they
could be detectable with GLAST and ground-based 
Cherenkov telescopes \cite{tor04}.  In the 
inner starburst regions of these sources, interaction of cosmic 
rays with the strong stellar winds  would produce 
GeV and TeV radiation \cite{rt03}. These processes will
also generate neutrinos, though it is unlikely that they
will  be detected with IceCube or a Northern Hemisphere km-scale
neutrino telescope, as these detectors have a sensitivity comparable
to EGRET in terms of fluence\footnote{A fuller discussion of 
neutrino sources will be given in my Madison proceedings for
TeV/Particle Astrophysics II.}.  

Torres (2004) \cite{tor04a} developed a detailed model of 
the nonthermal cosmic-ray production from the ultraluminous
infrared galaxy (ULIRG) Arp 220 at $\approx 72$ Mpc. ULIRGs 
are the result of merging galaxies that drive large quantities
of gas to the center of the system to form a dense
gas disk, trigger a starburst, and possibly fuel a buried 
AGN. Because of their
intense infrared emissions, Compton scattered
radiations from cosmic ray electrons on the IR photons 
could additionally enhance the \g-ray fluxes. In spite of 
its large distance, Arp 220 is potentially detectable with 
GLAST and the ground-based \g-ray telescopes \cite{tor04a},
because the dense clouds of target gas and increased cosmic ray 
confinement significantly increase
the brightness of ULIRGs in comparison to expectations from
a simple scaling to the Milky Way.

\paragraph{Clusters of Galaxies}

\begin{figure}
\centering
  \includegraphics[scale=0.4]{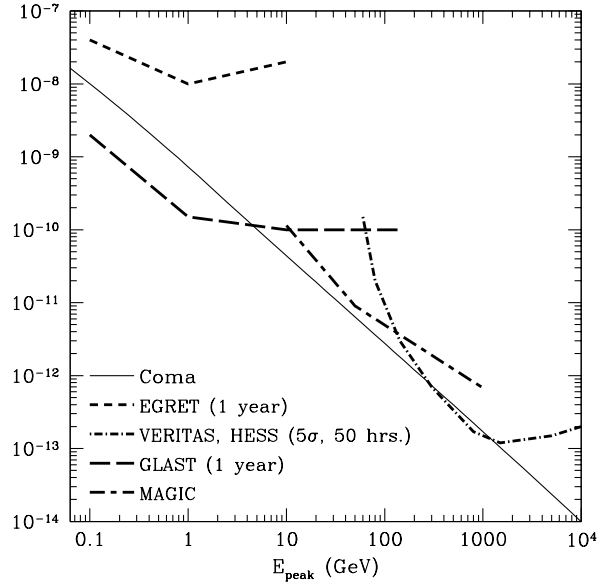}
\caption{Predicted $\gamma$-ray emission from the Coma cluster of galaxies
from the cluster merger shock model \cite{bd04}. The solid curve is the predicted photon
flux (in units of ph($> E_{\rm peak}$) cm$^{-2}$ s$^{-1}$).  Sensitivity
curves for EGRET, {\it MAGIC}, {\it GLAST}, and {\it VERITAS} and {\it HESS}
are shown.  The EGRET limits are for 2 weeks in the pointing mode, and the
{\it GLAST} limits are for 1 year in its scanning mode.  The quoted {\it
VERITAS}, {\it MAGIC} and {\it HESS} point-source sensitivities are for 50
hour, $5\sigma$ observations \cite{wee02}.}
\label{fig:2}      
\end{figure}

Nonthermal radiation will accompany 
the formation of collisionless shocks by merging clusters of galaxies
during the merger of dark matter halos in the standard model 
for the growth of structure in a $\Lambda$CDM universe \cite{bbu00}. The available 
energy in the merger between a cluster of mass $M_1 = 10^{15}M_{15} M_\odot$
and a smaller cluster of mass $M_2 = 10^{14}M_{14} M_\odot$, initially
separated by a distance of $r_1 = r_{Mpc}$ Mpc, is
\begin{equation}
{\cal E} \approx {GM_1 M_2\over r} \approx {8\times 10^{63}\over r_{\rm
Mpc}}\; M_{15}M_{14}\; {\rm ergs}\;.
\label{calE}
\end{equation}
If only a very tiny fraction of this energy is dissipated in the form of nonthermal cosmic rays
protons, then $\gtrsim 10^{60}$ ergs of cosmic rays, which are effectively 
trapped in the cluster even for a weak, $\sim 0.1\mu$G,  magnetic field, will be dissipated
on the mean timescale for a nuclear collision with cluster
gas. The thermal X-ray bremsstrahlung cluster emission shows that the thermal
cluster matter density is $n_{th} \approx 10^{-3}$ cm$^{-3}$, so that the characteristic
timescale for nuclear interactions is $t_{pp} \approx (n_{th}\sigma_{pp}c)^{-1} 
\cong 10^{18}$ s. Hence the bolometric
nonthermal cluster luminosity from pion producing interactions is
$L_{pp} \sim 10^{42}$ ergs s$^{-1}$.

Fig.\ 2 shows Berrington's calculations \cite{bd04} of 
the predicted $\gamma$-ray emission from the
Coma cluster of galaxies, 
using parameters appropriate to the
recent merger that has taken place in the Coma cluster 
environment ($d \cong 100$ Mpc). Point-source sensitivity limits for VERITAS
(which is comparable to the 
HESS sensitivity) are taken from Ref.\ \cite{wee02}, though the sensitivity may be degraded by Coma's 
angular extent \cite{gb04}.  The predicted
$\gamma$-ray emission falls below the EGRET sensitivity curve and the measured
2$\sigma$ upper limit of $3.81\times 10^{-8}$ ph($>100$ MeV) cm$^{-2}$
s$^{-1}$ \cite{rei03}.  The results from the merging cluster model show
 that GLAST will
significantly detect the non-thermal $\gamma$-rays from Coma to energies of
several GeV.  Furthermore, VERITAS could have a high confidence
($\gtrsim 5\sigma$) detection of Coma (at declination $+38^\circ$), depending on detail on the fraction of
energy going into shocked cosmic-ray protons, the nonthermal nature of Coma's hard X-ray spectrum,
and the amount of
nonthermal proton energy left over from previous merger events.  

\begin{figure}
\vskip-2.0in
\centering
  \includegraphics[scale=0.43]{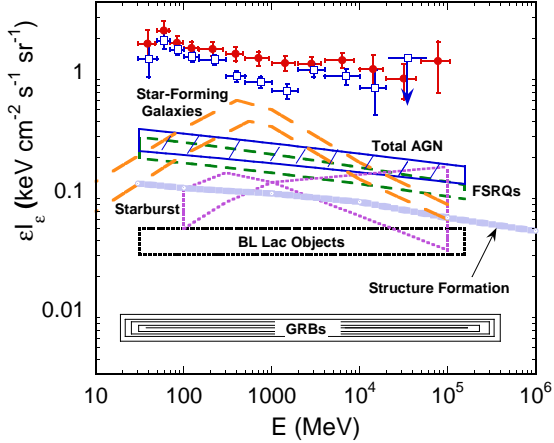}
%\vskip0.3in
\caption{Diffuse extragalactic $\gamma$-ray background, from 
analyses of EGRET data \cite{sre98,smr00}, compared to
model calculations of the contributions to the EGRB 
for FSRQ and BL blazars, and total AGNs \cite{der06}, star-forming galaxies \cite{pf02}, 
starburst galaxies \cite{tqw06}, structure 
shocks in clusters of galaxies \cite{kes03}, and all long-duration GRBs, 
including those detected as such (estimated herein). Pulsar 
contribution at 1 GeV is $\approx 20$\% of star-forming galaxy estimates. }
\label{fig:3}      
\end{figure}

In addition to merger shocks, high Mach number accretion shocks at the periphery of 
a forming cluster can accelerate nonthermal particles \cite{ryu03}.
The diffuse $\gamma$-ray background formed by intergalactic structure 
formation shocks from the calculations of Keshet et al.\ (2003) \cite{kes03}
is shown in Fig.\ 3.
Gamma-ray emission has not yet been convincingly detected from
clusters of galaxies \cite{rei03}, but calculations like this indicate that 
clusters of galaxies are likely to be the next established class of 
extragalactic sources of high-energy radiation. 

\paragraph{Diffuse Intensity from Star-Forming Galaxies}

We can use eq.\ (\ref{eIeiso}) to calculate 
the diffuse intensity from star-forming galaxies by normalizing
to the density and nonthermal \g-ray luminosity of $L_*$ galaxies
like the Milky Way. Fits of galaxy surveys 
to the Schechter luminosity function imply 
that the density of $L_*$ galaxies is $n_* = 0.016h^3$ Mpc$^{-3} \approx
1/(170$ Mpc$^{3}$) \cite{bm98}. Employing a mono-luminosity 
galaxy luminosity function, 
$\e_*^2 q_*(\e_*;z)$ $ = n_*\Sigma_*(z) \e_* L_*(\e_*)$, and eq.\ (\ref{eIeiso})
becomes, using eq.\ (\ref{estarLMW}),
$$\e I_\e^{*} = {c\over 4\pi }\;\int_0^\infty dz\; |{dt_*\over dz}|\;{\e_*^2 q_{*}(\e_*;z)
\over 1+z}\approx 4\times 10^{-7}\;\times$$
\begin{equation}
\big({\e\over 200}\big)^{-0.4}\int_0^\infty dz\;{\Sigma_*(z) (1+z)^{-2.4}\over
\sqrt{\Omega_m(1+z)^3 +\Omega_\Lambda}}\;{{\rm GeV}\over {\rm cm}^2~{\rm s}~{\rm sr}}\;.
\label{eIestar}
\end{equation}

Using the star formation rate function of Ref.\ \cite{hb06}, the integral
in eq.\ (\ref{eIestar}) is easily performed to give a value of $2.14$, so 
that $\e I_\e^{*} \cong 8.7\times 10^{-7} [E_\gamma/({\rm 100~MeV})]^{-0.4}$ GeV cm$^{-2}$
s$^{-1}$ sr$^{-1}$. This result is in good agreement with the more detailed 
treatment of the ``guaranteed \g-ray background" by Pavlidou and Fields (2002)
\cite{pf02}, plotted in Fig.\ 3.
The upper curve is scaled to a dust-corrected
star formation rate, and the lower curve stops the integration at $z = 1$.
Also shown is the intensity of 
starburst galaxies  estimated from a radio/FIR correlation \cite{tqw06}. 

\paragraph{Extragalactic Pulsar Emissions}

Using SAS-2 data
for the diffuse galactic $\gamma$-ray emission and EGRET
data for pulsars,  the analysis of Ref.\ \cite{sd96} shows
that the total $\gtrsim 100$ MeV flux of diffuse radiation from 
the Milky Way is $\approx 1.5\times 10^{-7}$ ergs cm$^{-1}$ s$^{-1}$, 
and the combined flux of the 6 brightest EGRET pulsars is 
$\approx 1.35\times 10^{-8}$  ergs cm$^{-1}$ s$^{-1}$. 
The modeling in that paper shows that the superposition 
of diffuse fluxes of unresolved pulsars is at the level
of $\approx 1.2\times 10^{-8}$ ergs cm$^{-1}$ s$^{-1}$. 
Thus total pulsar emissions make up as much as 20\% of the 
total galactic $\gamma$-ray flux in star-forming galaxies
like the Milky Way. 

The apparently diffuse emissions from pulsars in galaxies
throughout the universe is then, to first order, at the level
of $\approx 20$\% and proportional to the star formation 
history of the universe. A number of important effects must be 
considered for more accurate estimates of the extragalactic 
diffuse pulsar flux at different $\gamma$-ray
energies, most obviously being the harder pulsar 
spectrum (compared to the cosmic-ray induced emissions)
at energies up to the pulsar cutoff energies between $\approx
1$ -- 100 GeV \cite{tho97,str06}. Of great interest is to accurately measure
the high-energy pulsar spectral cutoffs with GLAST, which can be included in 
a more complete model for the pulsar contribution to 
the diffuse galactic background. Unfortunately, GLAST 
would not be sensitive to detect Milky-Way like 
$\gamma$-ray pulsars from 
nearby galaxies. Placing the Crab pulsar at 1 Mpc
would yield a $\gtrsim 100$ MeV apparent isotropic flux
$\lesssim 10^{-14}$ ergs cm$^{-2}$ s$^{-1}$.

\subsection{\g-rays and \n from Beamed Sources}

The evidence from EGRET and Whipple shows that beamed GRBs
and blazars are the brightest extragalactic high-energy \g-ray sources, 
and that isotropically emitting sources will be difficult
to detect except in a few cases, as just demonstrated. 

\paragraph{Microquasars}
One class of beamed source that has not yet been 
detected from beyond the Galaxy is the microquasar class, even 
though some 
of the ultraluminous X-ray sources seen in nearby 
galaxies could be microquasars with their jets oriented
towards us  \cite{gak02}. Presently only 
high-mass microquasars are known sources of GeV and TeV radiation, and the 
evidence of associations of $\gamma$-ray sources with low-mass systems is weak. 
The established cases of
\g-ray emitting microquasars are 
LS 5039, associated with an unidentified EGRET source
\cite{par00}
and unambiguously detected with HESS \cite{aha05}, and LSI +61 303,
whose orbital modulation has been recently demonstrated
with the MAGIC telescope \cite{alb06}. 

Models for microquasars do not require
large bulk Lorentz factors $\Gamma$ of the plasma outflow in 
microquasar jets \cite{brp06}, and superluminal motion observations generally
reveal microquasar Doppler factors $\lesssim 2$. 
The small sample leaves open the possibility that 
$\Gamma$ could exceed a few, which would make possible 
the detection of microquasars from distant galaxies. 
Using eq.\ (\ref{intep0}) and the threshold expression
$\delta_{\rm D}^q \ell_e^\prime \e^{\alpha_\nu}/d^2 \gtrsim 
f_\e \cong 10^{-12}$ ergs cm$^{-2}$ s$^{-1}$, we find that 
for a flat $\nu F_\nu$
spectrum ($\alpha_\nu = 2$) that $\delta_{\rm D}^q\gtrsim 6 (d/{\rm Mpc})^2$ for 
a Chandrasekhar mass compact object and a bolometric factor 
$\lambda_b = 10$. Thus, Doppler factors of only a few or greater
are needed in order to detect microquasars from galaxies at distances
$\gtrsim 1$ Mpc.

\paragraph{Gamma Ray Bursts}

A detailed treatment of the statistics of GRBs is given in 
Ref.\ \cite{ld06}. We can use those results and eq.\ (\ref{eIeGRB}) 
to estimate the diffuse \g-ray background from GRBs. To simplify
the results, we consider all GRBs, including those above 
threshold, and approximate the GRB spectrum as 
a flat $\nu F_\nu$ spectrum to the highest \g-ray energies.
The diffuse \g-ray background intensity is then given by
\begin{equation}
\e I_\e^{GRB} \cong {c \dot n_{GRB}{\cal E}_{*\gamma} \over 4\pi \lambda_b H_0}
\int_0^\infty dz\; {\Sigma_{GRB}(z)\over (1+z)^2} \;.
\label{eIeGRB1}
\end{equation}
For a local GRB event rate $\dot n_{GRB} = 10\dot n_{10}$ Gpc$^{-3}$ yr$^{-1}$, 
\begin{equation}
\e I_\e^{GRB} \cong {3\times 10^{-9} \dot n_{10} \over \lambda_{10}}\int_0^\infty dz\; {(1+z)^{-2}\Sigma_{GRB}(z)\over 
\sqrt{\Omega_m(1+z)^3 +\Omega_\Lambda}},
\label{eIeGRB1}
\end{equation}
in units of GeV ${\rm cm}^{-2}~{\rm s}^{-1}~{\rm sr}^{-1}$.

Using star formation rates 5 and 6 that allow the redshift and jet opening 
angle distributions to be fit \cite{ld06} gives  
the diffuse intensity from GRBs shown in Fig.\ 3 which
 is a small fraction of the diffuse $\gamma$-ray background.

\paragraph{Blazars}

EGRET and GLAST data on blazars can be analyzed 
with models that jointly fit
the redshift and flux size distribution and predict the level of 
the  extragalactic $\gamma$-ray background (EGRB) \cite{mp00,der06}.
The diffuse intensities of unresolved FSRQ (dashed) 
and BL Lac (dotted) blazars, and the total AGN contribution (shaded),
 are shown in Fig.\ 3 \cite{der06}. The ranges
correspond to sensitivities  $\phi_{-8} = 25$ and $\phi_{-8} = 12.5$.
 The BL Lac objects and FSRQs, 
including emissions from misaligned radio galaxies, contribute
at the $\sim 2$ -- 4\% and $\sim 10$ -- 15\% levels, 
respectively, to the total EGRB \cite{sre98}
near 1 GeV. 

The sum of the different contributions in Fig.\ 3 at $\approx 1$ GeV is at about the level 
of the total diffuse extragalactic
$\gamma$-ray emissions measured with EGRET \cite{sre98}. Soft blazar sources, and softer than modeled diffuse 
cosmic ray emissions from normal galaxies, could account for the residual emissions 
between $\approx 50$ MeV -- 1 GeV. New hard $\gamma$-ray source populations are apparently 
required at $\gtrsim 10$ GeV, which would include cascade emission from UHE electromagnetic
cascades with photons of the EBL.

\section{Summary and Conclusions}

It will be of considerable interest when GLAST or a TeV telescope
detects M31 or another galaxy of the local group, or
a starburst or IR luminous galaxy. 
The measured flux will give a valuable check on the efficiency of cosmic
ray production as a function of galactic star formation 
activity, and will provide a further normalization,
after the Milky Way and the LMC, of the contribution of star forming 
galaxies to the $\gamma$-ray background.

GLAST will provide large statistical samples on at least two source classes:
GRBs and blazars.  There are good reasons to think that GLAST will
detect star forming galaxies and clusters of galaxies. Although the
sensitivity to extended sources is degraded in TeV telescopes, the Coma
cluster is near the threshold for detection, depending on its nonthermal
X-ray spectrum, but M31 would require 
fortunate spectral and spatial emissions to be detected.

For GRBs and blazars, there are already fundamentally interesting
questions about the evolution of the source rate densities and 
change in properties of relativistic jet sources through cosmic time. 
For $\gamma$-ray emitting BL Lac objects, 
positive source evolution (more sources at late times) and negative
luminosity evolution (sources brighter in the past) explains the statistical 
distributions from EGRET \cite{der06}. Detection of blazars to 
high redshift, $z\gg 5$, is expected with GLAST.
It will be interesting to see if GLAST LAT-detected
GRBs are peculiar in their properties compared 
to long-duration GRBs found by burst detectors with
$\sim 100$ keV triggers. 

The EGRB is probably a composite of many source
classes (Fig.\ 3), which can best be established by identifying 
individual sources with better sensitivity detectors. The contribution of beamed to 
unbeamed sources, after subtracting
known sources, is limited by the $\gamma$-ray statistical excursions of the 
EGRB measured with GLAST. GLAST will monitor blazar and GRB flaring with its $\sim 2$ sr field of view,
and rapidly slewing instruments like MAGIC may soon discover the first VHE, $\gg 10$ GeV, GRB. 
But a wide field-of-view ground-based $\gamma$-ray telescope, like HAWC or the 
next generation TeV telescopes, will have the best chance
to monitor TeV $\gamma$-ray transients for follow-up observations. 
Improved statistical
analyses of $\gamma$-ray and particle astronomy projects will tell us the composition
of the unresolved residual $\gamma$-ray emission, and if there are room for more source classes, such as 
dark matter emissions, anomalous microquasars or odd classes of $\gamma$-ray 
emitting objects yet to be discovered.

%%%%%%%%%%%%%%%%%%%%%%%%%%%%%%%%%%%%%%%%%%%%%%%%%%%%%%%%%%%%%%%%%%%%%%%%%%%%%%%
%%%%%%%%%%%%%%%%%%%%%%%%%%%%%%%%%%%%%%%%%%%%%%%%%%%%%%%%%%%%%%%%%%%%%%%%%%%%%%%

\begin{acknowledgements}
I would like to thank the organizers, Josep M.\ Paredes, Olaf Reimer,
and  Diego F.\ Torres, for the kind invitation to speak at this conference,
and for the opportunity to visit the beautiful city of Barcelona.
I would also like to thank A.\ Atoyan, T.\ Le, and V.\
Vassiliev for discussions, and to Dr. Vassiliev for 
correcting an error in the draft version. This work is supported by the
Office of Naval Research and a GLAST Interdisciplinary
Scientist grant.
\end{acknowledgements}

\end{document}